\documentclass[journal]{IEEEtran}
\usepackage{xcolor,soul,framed} 

\colorlet{shadecolor}{yellow}
\usepackage[pdftex]{graphicx}
\graphicspath{{../pdf/}{../jpeg/}}
\DeclareGraphicsExtensions{.pdf,.jpeg,.png}

\usepackage[cmex10]{amsmath}
\usepackage{array}
\usepackage{mdwmath}
\usepackage{mdwtab}
\usepackage{eqparbox}
\usepackage{url}
\usepackage{listings}
\usepackage{multicol}

\begin{document}
\bstctlcite{IEEEexample:BSTcontrol}
    \title{Direct On-Wafer Measurements of Noise Parameters\\ in C- and X-bands at $T=4$~K}
  \author{Daniil~Frolov,~Jean-Olivier~Plouchart,~Utku~Soylu

  \thanks{Daniil Frolov (e-mail: drf@ibm.com), Jean-Olivier Plouchart (e-mail: plouchar@us.ibm.com), Utku Soylu (e-mail: utkusoylu@ibm.com) are with IBM Quantum, Thomas J. Watson Research Center, Yorktown Heights, New York, USA. 
  
  This work has been submitted for possible publication. Copyright may be transferred without notice, after which this version may no longer be accessible.}
  }
   
\maketitle

\begin{abstract}
\mdseries
This paper describes the setup and the results of the direct on-wafer measurements of a FET noise parameters obtained with a source-pull method at temperatures down to T~=~4~K and in the 5~--~12~GHz frequency range. The setup consists of a cryostat with wafer probes, two reflectometers, a programmable impedance generator, wideband isolators and bias tees and low noise preamplifier, all cooled to cryogenic temperatures, allowing to perform a full vector error-corrected wafer-level measurements of the discrete transistors and amplifier dies. The setup and its calibration procedure are designed in a such way that allows simultaneous calibration, S-parameters, noise parameters and I-V curve measurements of several FETs all in one cooldown. Using the described setup we perform first measurements of 14nm FinFETs and also measure noise parameters of an LNA based on these FETs. Resulting noise temperature values are compared against those obtained using independent and alternative measurement techniques.
\end{abstract}


%
\IEEEpeerreviewmaketitle


\section{Introduction}

The modeling and characterization of transistors and LNAs for cryogenic applications present significant technical challenges and have been addressed in a number of prior works \cite{russel_paper}, \cite{cold_attn}, \cite{onchip_paper}, \cite{Kelly}, \cite{bardin_paper}. Historically, these applications were predominantly associated with radio astronomy, space, and physics-oriented research technologies. As a result, cryogenic process design kit (PDK) development remained a niche activity, often focused on a limited set of specialized devices. 

Commercial superconducting quantum computing systems, such as IBM Starling and Blue Jay machines \cite{ibm}, which are currently under active development, are expected to operate with tens of thousands of physical qubits. This scale requires readout chains incorporating a variety of low-noise transistor-based and quantum-limited superconducting amplifiers operating at the lower end of the X-band. This readout application, together with related applications such as cryogenic CMOS circuits for qubit control and state discrimination, is therefore creating a growing demand for cryogenic PDKs. The development of such PDKs is impossible without wafer-level characterization tools that enable chip sorting, binning, and the generation of large datasets for yield analysis and statistical evaluation.

Previously, we demonstrated two scalar noise measurement methodologies for on-wafer IC characterization that utilize both one-step and two-step cooldown procedures \cite{arftg_paper}. In recent years, impedance generators capable of operating at temperatures as low as 4~K have become available, largely due to the work of Leo Belostotski and his colleagues \cite{leo_patent}, \cite{Belostotski}. These impedance generators are now commercially available from Maury Microwave \cite{maury}. However, to the best of the authors' knowledge, cryogenic noise parameter measurements using these impedance generators have, until recently, been performed only on packaged LNAs, and not on wafer-level devices, or, more importantly, on individual discrete transistors on wafer. Among several challenges that make these measurements particularly complex and labor-intensive is the requirement to perform multiple cryostat cooldowns to obtain all necessary calibration data. This process can take more than a day to complete for a single LNA. 

This paper describes a novel measurement system together with a corresponding calibration scheme that enables cryogenic noise parameter measurements using source-pull and cold-source techniques. Compared to previously reported setups, the proposed approach offers several significant advantages:
\begin{enumerate}
\item Unlike earlier systems \cite{russel_paper}, \cite{cold_attn}, \cite{onchip_paper}, the described setup enables direct \textit{on-wafer} noise parameter measurements \textit{and} simultaneous S-parameter measurements of cryogenic DUTs with full vector error correction, over a frequency range with an arbitrary frequency step defined by the operator;

\item The DUT may be either a high-gain amplifier or a low-gain single discrete transistor, without degradation of the signal-to-noise ratio imposed by the 300~K noise floor limit of a room-temperature noise analyzer;

\item In the proposed setup, S-parameter calibration, noise calibration, and DUT measurements are all performed within a single cryostat cooldown. This capability allows multiple devices on a single wafer to be characterized in one run, significantly reducing test time and enabling automated, mass-production-level testing.
\end{enumerate}
\begin{figure}[h]
  \begin{center}
  \includegraphics[width=3.5in, clip, trim=20pt 20pt 20pt 20pt]{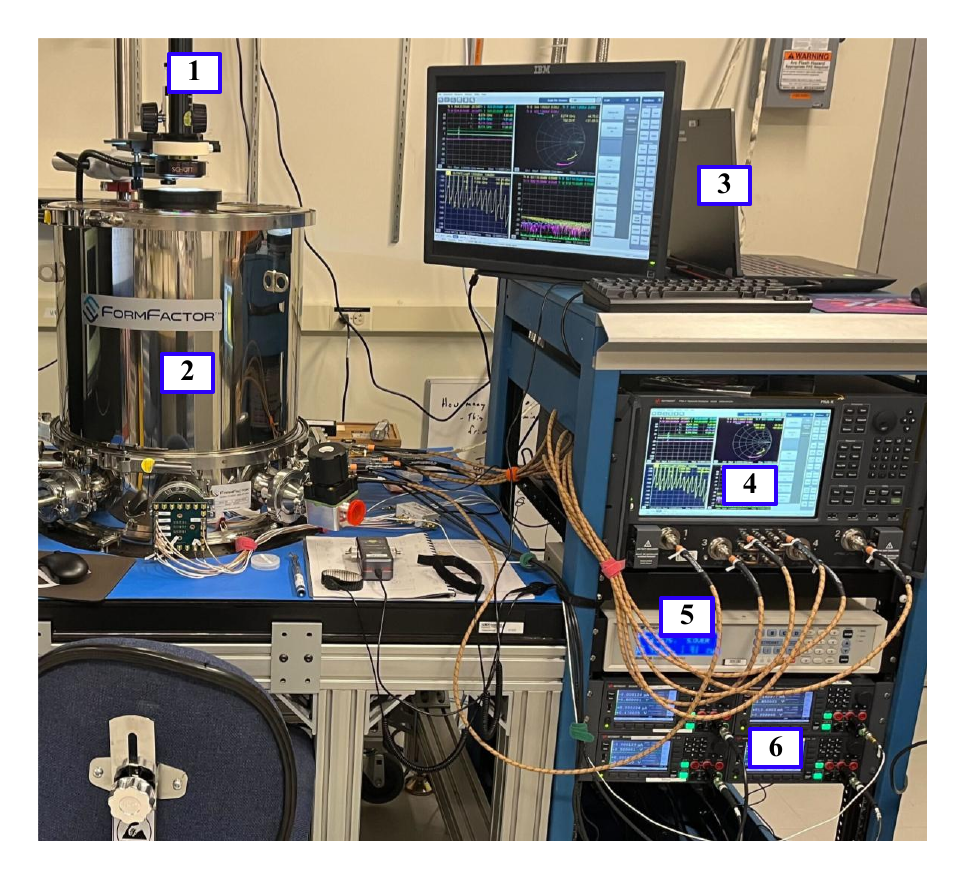}\\
  \caption{Noise parameter measurement stand consists of microscope~1, cryostat~2, control computer~3, vector network analyzer (VNA)~4, temperature readout~5, source-measure units (SMU)~6 for DC-biasing.}\label{setup_main_view}
  \end{center}
\end{figure}

Section~II of this paper provides a detailed technical description of the cryogenic setup and configuration of the external room-temperature instrumentation. Section~III discusses the calibration procedures of the setup and outlines the features and limitations of the measurements. Section~IV describes the 14~nm FinFET transistors under study, followed by their noise parameter measurements. Section~V presents the LNA based on these FinFETs, along with measurements of its S-parameters and noise parameters obtained using two independent methods:
\begin{enumerate}
\item using the source-pull and cold-source methods with vector error correction in the described noise parameter measurement setup;
\item using the cold-attenuator method and thru scalar normalization, as described in \cite{arftg_paper}.
\end{enumerate}
Results are discussed in Section VI, followed by a Conclusion. Similar to \cite{russel_paper}, in this work we utilize the following noise parameter set for a given DUT:
\begin{equation}
T_{e} = T_{min} + 4NT_0 \frac{|\Gamma_s-\Gamma_{opt}|^2}{(1-|\Gamma_s|^2)(1-|\Gamma_{opt}|^2)} \label{eq:noise_param}
\end{equation}
where: $T_{min}$ is the minimum effective noise temperature of the device; $N$ is Lange invariant parameter \cite{lange_N};
\mbox{$\Gamma_{opt}=(R_{opt}+jX_{opt}-Z_0)/(R_{opt}+jX_{opt}+Z_0)$}, with $R_{opt}$ and $X_{opt}$ corresponding to the optimal resistance and reactance of the source connected to the DUT which allows to achieve condition when $T_e=T_{min}$; $T_0=290$~K; $Z_0=50\: \Omega$.

\section{Instrumentation}
\subsection{General Description}
\begin{figure}[h]
  \begin{center}
  \includegraphics[width=3.5in,  clip, trim=20pt 20pt 20pt 20pt]{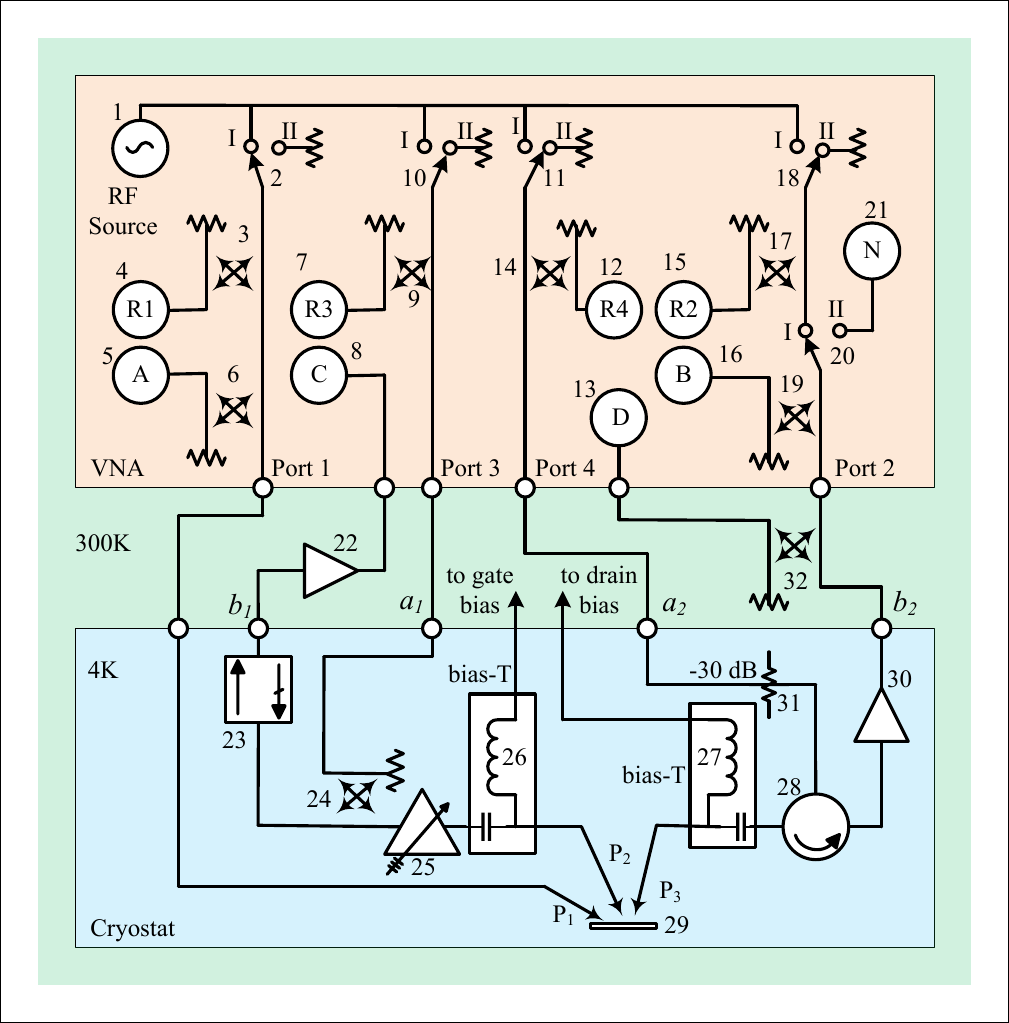}\\
  \caption{Electrical block-diagram of the measurement instruments and hardware. Two major components are the VNA (Keysight 5242B) and the cryo wafer prober (Form Factor), in between there are minor room temperature components amplifier~22 (Mini-Circuits) and directional coupler~32. 
  The VNA equivalent circuit includes: RF source~1; terminated switches~2,10,11,18; SPDT switch~20; heterodyne receivers~R1...R4, A,B,C,D; noise receiver~21; and directional couplers~3,6,9,14,17,19.
  The RF hardware inside the cryostat includes: 4-12~GHz triple junction ferrite isolator~23, 3.4-12.3~GHz circulator~28, 0.3-14~GHz low noise amplifier (Low noise factory)~30, 30~dB directional coupler~24, 0.2-18~GHz bias-T~27 (Quantum Microwave), impedance tuner~25 with built in bias-T~26 (Maury Microwave), attenuator~31, wafer probes $P_1$, $P_2$, $P_3$ (GGB Industries) and sample holder~29.}\label{block-diagram}
  \end{center}
\end{figure}

\begin{figure}[h]
  \begin{center}
  \includegraphics[width=3.5in, clip, trim=20pt 20pt 20pt 20pt]{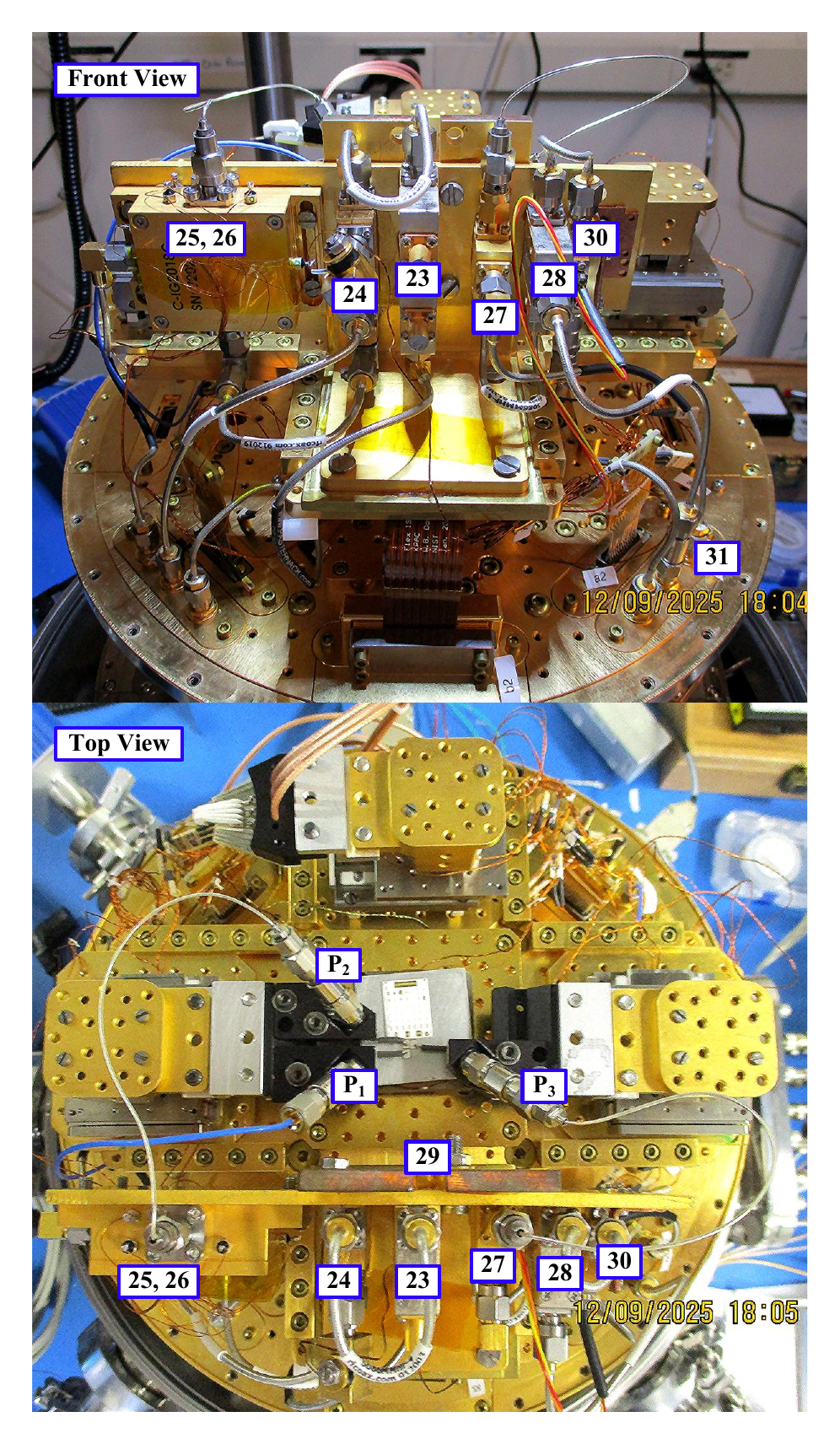}\\
  \caption{Cryogenic instrumentation mounted on the 4K plate of the refrigerator: front view and top view (as seen by the microscope), with the same numbering as in block diagram in fig.\ref{block-diagram}. Probes $P_1$, $P_2$, $P_3$ are moved by Attocube piezo-based drives for cryogenic environments.}\label{inside}
  \end{center}
\end{figure}

The measurement setup consists of commercially available instruments and hardware and is shown in Fig.~\ref{setup_main_view}. The electrical block diagram of the system is presented in Fig.~\ref{block-diagram}. The setup comprises two major subsystems that enable simultaneous, direct \textit{on-wafer} noise parameter and S-parameter measurements: a four-port vector network analyzer with a built-in noise receiver, and a cryogen-free 4~K refrigerator equipped with cryogenic microwave instrumentation, as illustrated in Fig.~\ref{inside}. 

The primary room-temperature instrument in the setup is the network analyzer. In this work, we used a Keysight PNA-X model N5242B; however, an equivalent ZNA26 model from Rohde \& Schwarz or a similar instrument may be used with comparable results. The main requirements are that the analyzer provides four ports, direct access to the internal receivers, and a built-in noise receiver. While the use of a simpler VNA is possible, it would require multiple external switches and a separate noise figure analyzer, significantly complicating the calibration procedure. The electrical diagram of the network analyzer shown in Fig.~\ref{block-diagram} is a generic equivalent and includes a reduced number of components, limited to those essential for understanding the measurement and calibration concepts.

Noise parameter measurements are performed by implementing a combination of cold-source and source-pull methods \cite{vna_book} directly at cryogenic temperatures using a cryogenic impedance generator (referred to as a ``tuner'' in this work)~25. This tuner is based on the original work by M.~Himmelfarb and L.~Belostotski \cite{leo_patent} and is currently commercially available from Maury Microwave \cite{maury}. The tuner operates in four discrete states, labeled 1--4. Only in state~1 does it function as a two-port device with relatively well-matched input and output ports and low insertion loss. In the remaining three states, the tuner effectively becomes a one-port network presenting three fixed output impedances over the frequency range of interest, as will be discussed later.

To isolate the low-noise cryogenic instrumentation from room-temperature noise generated by the network analyzer, a ferrite isolator~23, a high-coupling-loss directional coupler~24, and an attenuator~31 are employed. Although the tuner itself is designed to operate over the 2--18~GHz frequency range, the effective measurement bandwidth of the setup is primarily limited by the ferrite isolator and the on-wafer TRL calibration to the 5--12~GHz range. Components~23, 24, 28, and~31 not only isolate the DUT from room-temperature noise, but also collectively function as directional elements that separate the forward waves $a_1$, $a_2$ from the reflected waves $b_1$, $b_2$ directly in the cryogenic environment. Together, they form an external cryogenic S-parameter test set that is connected to the network analyzer through the direct receiver access ports on the front panel.

Another critical component enabling noise parameter measurements of individual transistors in this setup is the preamplifier~30. For example, a 4~K noise power spectral density (PSD) of $-193$~dBm/Hz from a cold 50~$\Omega$ termination, amplified by a transistor with a typical gain of $+10$~dB and a low effective noise temperature $T_e$, results in an output noise PSD of approximately $-183$~dBm/Hz. This level remains well below the room-temperature instrumentation noise floor of $-174$~dBm/Hz, rendering accurate measurement impractical. The preamplifier~30 raises the output noise level significantly above the room-temperature noise floor, enabling measurements of transistors with very low gain. However, accurate and careful calibration of this preamplifier is essential for reliable noise parameter extraction.

\subsection{Principles of Operation}
The setup shown in Fig.~\ref{block-diagram} operates in two distinct modes: a) S-parameter measurement and b) noise parameter measurement. Both modes require calibration at both room temperature and cryogenic temperature, as discussed in the Calibration section.

\subsubsection{S-parameter mode}
In this mode, the S-parameters of any two-port DUT, or of two single-port DUTs connected to probes $P_2$ and $P_3$, can be measured. The DUT may be a passive device, such as an attenuator or an inductor, or an active device, such as a transistor. In the latter case, the required biasing can be supplied to the probes through bias-Ts~26 and~27. Alternatively, the DUT may be a fully featured multi-stage LNA with external DC probes used for power delivery.
Measurements in this mode are performed using VNA ports~3 and~4. Switches~2 and~18 are fixed in position~II for all S-parameter measurements. During the forward scan ($S_{11}$, $S_{21}$), switches~10 and~11 are set to positions~I and~II, respectively, while for the reverse scan ($S_{22}$, $S_{12}$), the switch positions are reversed.

During the forward scan, the incident wave $a_1$ is generated by RF source~1 and routed through switch~10, coupler~9, and port~3 to the cryogenic directional coupler~24. Within this coupler, $a_1$ is strongly attenuated by the coupling loss and directed toward probe $P_2$ via tuner~25 and bias-T~26. The attenuation provided by coupler~24 is necessary to suppress the room-temperature instrumentation noise floor of $-174$~dBm/Hz down to the cryogenic instrumentation noise floor of $-193$~dBm/Hz.

The wave $b_1$ reflected from the DUT propagates back through probe $P_2$, bias-T~26, and tuner~25 into directional coupler~24, where it is routed through the primary low-loss channel toward the network analyzer receiver ``C''~8 via isolator~23 and amplifier~22. In this configuration, $S_{11}$ is measured as $S_{11} = b_1/a_1 = C/R_3$. The wave $b_2$ transmitted through the DUT propagates forward through probe $P_3$, bias-T~27, circulator~28, and preamplifier~30, and is then directed to receiver ``D''~13 through the secondary channel of directional coupler~32. Accordingly, $S_{21}$ is measured as $S_{21} = b_2/a_1 = D/R_3$.

It can be observed that the forward propagation path of wave $a_1$ experiences high attenuation, whereas the return path of the reflected wave $b_1$ exhibits minimal attenuation. This asymmetric behavior is achieved primarily through the use of directional coupler~24 and isolator~23. The unidirectional propagation characteristic of isolator~23 effectively isolates the DUT from room-temperature noise while preserving a low-loss path for the reflected wave $b_1$ within the directional coupler~24.

During the reverse scan, the incident wave $a_2$ is generated by RF source~1 and routed through switch~11, coupler~14, and port~4 to the cryogenic attenuator~31. In this attenuator, $a_2$ is again significantly attenuated and directed toward probe $P_3$ via circulator~28 and bias-T~27. The wave $b_2$ reflected from the DUT propagates back through probe $P_3$ and bias-T~27 into circulator~28, where it is routed toward network analyzer receiver ``D''~13 through preamplifier~30 and the secondary channel of directional coupler~32. In this case, $S_{22}$ is measured as $S_{22} = b_2/a_2 = D/R_4$. 

The wave $b_1$ transmitted through the DUT in the reverse direction propagates through probe $P_2$, bias-T~26, and tuner~25 into directional coupler~24, and from there to network analyzer receiver ``C''~8 via isolator~23 and amplifier~22. Consequently, $S_{12}$ is measured as $S_{12} = b_1/a_2 = C/R_4$.

Amplifier~22 increases the sensitivity of receiver~C in the VNA; however, its primary purpose is to enable the calibration protocol described later in this paper.

\subsubsection{Noise parameter mode}
In this mode, the noise parameters of a two-port active DUT are measured. This procedure requires the measurement of four noise factors, $F_1$, $F_2$, $F_3$, and $F_4$, corresponding to four output impedance states, $Z_1$, $Z_2$, $Z_3$, and $Z_4$, of tuner~25. The noise factors are measured using the cold-source method \cite{vna_book} as follows.

After the S-parameters of the DUT are measured, switch~20 of the VNA is set to position~II, connecting the noise receiver ``N''~21 to the output of the DUT through the primary channel of directional coupler~32, cryogenic preamplifier~30, circulator~28, and bias-T~27. This configuration enables the measurement of the noise factors $F_i$ for each of the four tuner impedance states using the following equation:
\begin{equation}
F_i = 1 + \frac{1}{T_0} \Bigg[\frac{T_i}{|S_{21}|^2 M_i} - T_{amb}\Bigg]
\label{eq:noise_factor}
\end{equation}
where $S_{21}$ is the complex transmission coefficient of the DUT measured between probe tips $P_2$ and $P_3$; $T_{amb}$ is the tuner and the DUT physical temperature which are in thermal equilibrium; $T_i$ is the absolute available output noise temperature from the DUT for the specific tuner state measured directly at the probe tip $P_3$, and $M_i$ is the DUT mismatch term for each tuner state: 
\begin{equation}
M_i = \frac{1-|\Gamma_i|^2}{|1-\Gamma_i S_{11}|^2 \Big(1-\Big|S_{22} + \frac{S_{21}S_{12}\Gamma_i}{1-S_{11}\Gamma_i}\Big|^2 \Big)}  
\label{eq:match}
\end{equation}
where \mbox{$\Gamma_{i}=(Z_i-Z_0)/(Z_i+Z_0)$} complex reflection coefficient of the tuner output as seen by the DUT at the $P_2$ probe tip, $S_{11}$, is the complex reflection coefficient of the DUT at the $P_2$ probe tip, and $S_{22}$ at the probe tip $P_3$, while $S_{21},\: S_{12}$ are transmission coefficients between these probe tips.  

Noise parameters of the DUT can then be obtained from the following equations (see \cite{Belostotski}):
\begin{equation}
\begin{split}
F_{min} &= \mathrm{Re}(A+\sqrt{4BC-D^2}) \\
T_{min} &= T_0(F_{min}-1) \\
Y_{opt} &= \frac{1}{2B} \Big(\sqrt{4BC - D^2} - jD \Big) \\
N &= \mathrm{Re}(Y_{opt}) B \\
\Gamma_{opt} &=(Y_{opt}^{-1} - Z_0)/(Y_{opt}^{-1} + Z_0)
\end{split}
\label{eq:noise_param1}
\end{equation}
where coefficients $A$, $B$, $C$, $D$ can be calculated from the measured noise factors $F_1$, $F_2$, $F_3$, $F_4$ and output impedances $Z_i = R_i + jX_i$ of the tuner:
\begin{equation}
\label{eq:matrix}
\begin{pmatrix}
A\\
B\\
C\\
D\\
\end{pmatrix}
=
\begin{pmatrix}
1, & R_1^{-1}+ R_1^2X_1^{-2}, & R_1, & R_1 X_1^{-1} \\
1, & R_2^{-1}+ R_2^2X_2^{-2}, & R_2, & R_2 X_2^{-1} \\
1, & R_3^{-1}+ R_3^2X_3^{-2}, & R_3, & R_3 X_3^{-1} \\
1, & R_4^{-1}+ R_4^2X_4^{-2}, & R_4, & R_4 X_4^{-1} \\
\end{pmatrix}^{-1}
\begin{pmatrix}
F_1\\
F_2\\
F_3\\
F_4\\
\end{pmatrix}
\end{equation}

It can be seen that the calculation of noise parameters using (\ref{eq:noise_factor}) and (\ref{eq:matrix}) requires the S-parameter calibration plane to be moved to the probe tips $P_2$ and $P_3$. In addition, the noise PSD calibration plane must be located at probe tip $P_3$, and the tuner output impedances at probe tip $P_2$ must be known for all four tuner states. All of these requirements can be satisfied using the calibration procedures described below.

\section{Calibration Principles}
\begin{figure*}[h]
  \begin{center}
  \includegraphics[width=6in, clip, trim=15pt 15pt 15pt 15pt]{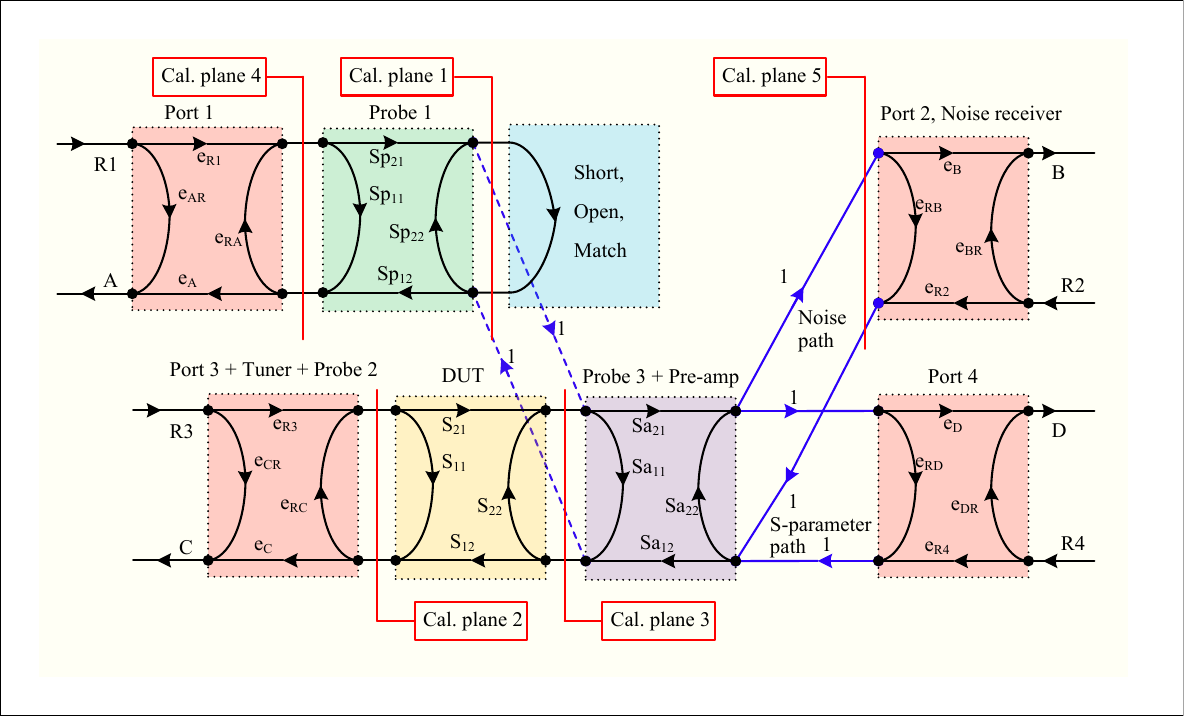}\\
  \caption{Signal graph of the system representing DUT, error boxes and five calibration planes required for calibration of 24 error terms.}\label{graph}
  \end{center}
  \vspace{-1.5em}
\end{figure*}
Calibration is a key step in noise parameter measurements. In our setup, both calibration and measurement are performed within a single cooldown, which significantly reduces the overall characterization time. The equivalent signal flow graph of the setup, shown in Fig.~\ref{graph}, includes five calibration planes required to de-embed the associated error boxes down to the probe tips. The DUT is located between planes~2 and~3 (probe tips $P_2$ and $P_3$). Depending on the type of measurement (noise parameter or S-parameter), the corresponding set of error boxes must be de-embedded. In total, 24 error terms must be de-embedded in order to obtain calibrated DUT S-parameters and noise parameters within a single cooldown.

\subsection{On-wafer S-parameter calibration}
As described previously, the four S-parameters of the DUT, $S_{11}$, $S_{22}$, $S_{21}$, and $S_{12}$, are measured using ports~3 and~4 of the VNA. Consequently, the DUT is bounded by two error boxes: \textit{``Port3+Tuner+Probe2''} on the input side and \textit{``Probe3+Preamp+Port4''} on the output side, as defined by calibration planes~2 and~3 in Fig.~\ref{graph}. Note that, on the output side, the \textit{``Probe3+Preamp''} and \textit{``Port4''} error boxes are combined into a single equivalent error box. This simplification is sufficient for S-parameter measurements, although it is not adequate for noise parameter measurements, as will be discussed later.

S-parameter calibration is therefore performed between planes~2 and~3 (probe tips $P_2$ and $P_3$) using the standard multiline TRL (thru--reflect--line) method with a CS-105 calibration substrate and with the tuner switched to its first state.

\subsection{Tuner characterization}
\begin{figure}[h]
  \begin{center}
  \includegraphics[width=3.5in, clip, trim=10pt 10pt 10pt 10pt]{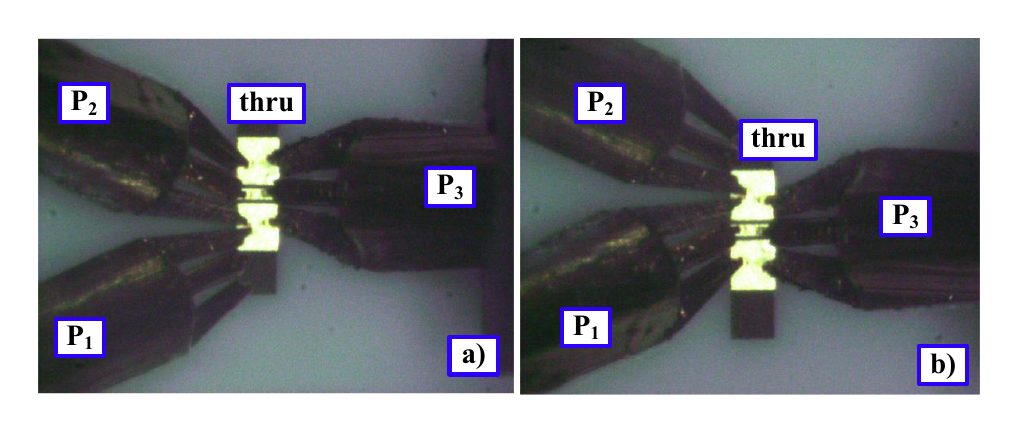}\\
  \caption{Characterization of the pre-amplifier. (a) preamp is connected to the tuner, (b) preamp is connected to the $P_1$ for S-parameter measurement}\label{thru}
  \end{center}
\end{figure}

The output impedances of the tuner, as seen by the DUT at probe tip $P_2$ in each of the four tuner states, are required for (\ref{eq:match}) and (\ref{eq:matrix}). These impedances are measured as follows. After completion of the S-parameter calibration, probes $P_2$ and $P_3$ remain landed on the thru standard of the calibration wafer, as shown in Fig.~\ref{thru}(a). This configuration directly connects calibration planes~2 and~3, such that the corrected reflection measured by port~4 of the VNA is $S_{22}=0\pm\delta$, where $\delta$ is the residual reflection uncertainty of the two-port calibration, typically $\delta \ll 0.01$, corresponding to better than $-40$~dB.
Because the tuner in state~1 is included in the S-parameter calibration procedure, its output impedance mismatch is fully removed during the correction process. However, if, \textit{after} calibration, the error terms $e_{CR}=e_{C}=0$ are enforced by disabling amplifier~22 and thereby forcing wave $b_1=0$, port~4 of the VNA directly measures the reflection at probe tip $P_2$ as seen by the DUT. This effectively yields a one-port network measurement, enabling characterization of the tuner output impedance $Z_i$ at probe tip $P_2$ for all four tuner states.

\begin{figure}[h]
  \begin{center}
  \includegraphics[width=3.5in, clip, trim=10pt 10pt 10pt 10pt]{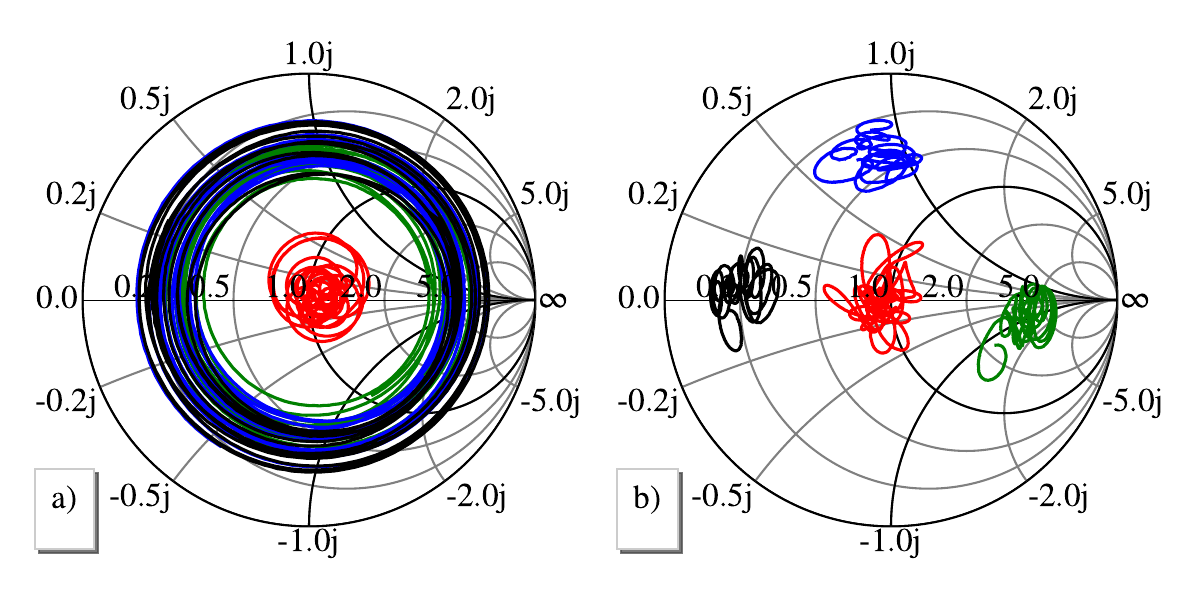}\\
  \caption{Measured probe $P_2$ source impedance as seen by the DUT (a) and with 1.923 ns delay subtracted (b): tuner output impedances $Z_1$ red (center), $Z_2$ green (right), $Z_3$ blue (top), $Z_4$ black (left).}\label{tuner_smith}
  \end{center}
\end{figure}

The results of this one-port measurement are shown in Fig.~\ref{tuner_smith}(a). As illustrated in Fig.~\ref{inside}, the tuner is connected to probe tip $P_2$ through a long cryogenic cable required to accommodate the motion range of the piezo-positioner. This cable, together with the electrical length of the probe itself, introduces an additional delay of 1.923~ns in the tuner impedance in our setup. The four tuner impedances over the 5--12~GHz frequency range, with this delay de-embedded, are shown in Fig.~\ref{tuner_smith}(b). Since the impedances cluster into four distinct regions, as required by the technique described in \cite{leo_patent}, this result confirms that the tuner operates as expected.
In fact it is not needed to know exactly how much delay the wiring adds and we are showing fig.\ref{tuner_smith} (b) only for illustrative purposes. For the noise parameter measurement only the actual impedance at the DUT input is important and it is directly measured in our setup, as seen in fig.\ref{tuner_smith} (a).     

\subsection{Pre-amplifier characterization}
The final step of the calibration procedure involves a complete characterization of the preamplifier S-parameters and noise parameters between probe tip $P_3$ and VNA port~2 (the noise receiver input).

\subsubsection{S-parameters characterization}
The S-parameters of the preamplifier are measured using an additional probe, $P_1$. To characterize this probe itself, a room-temperature calibration is first performed by directly connecting calibration planes~4 and~5 using a standard electronic calibrator (e-cal). A conventional room-temperature two-port calibration is then carried out between ports~1 and~2 of the VNA. This calibration is performed using the same frequency range and frequency step as the multiline TRL calibration previously executed in the cryostat.

The genders of the cable connectors between the VNA and the cryostat are selected such that no adapters are required when connecting calibration planes~4 and~5 either to the e-cal or to the cryostat. After completion of the port~1--port~2 calibration, the calibrated cables are reconnected to the cryostat. Subsequently, probe $P_1$ is landed on the Open, Short, and Match (OSM) standards of the calibration substrate inside the cryostat, and the corresponding reflection coefficients, $\Gamma_O$, $\Gamma_S$, and $\Gamma_M$, are measured by port~1 of the VNA.

Since the probe is a reciprocal device, $Sp_{21}$=$Sp_{12}$ its S-parameters then can be obtained using the equations \ref{eq:sol}:
\begin{equation}
   Sp_{11} = \text{EDF}, \: Sp_{22} = \text{ESF}, \: Sp_{21}=Sp_{12}=\sqrt{\text{ERF}} \label{eq:sol}
\end{equation}
where EDF, ESF and ERF are directivity, source match and reflection tracking error terms calculated from the measured values $\Gamma_O$, $\Gamma_S$, and $\Gamma_M$ using the OSM (also known as Short Open Load (SOL) one-port calibration matrix solution\footnote{OSM/SOL calibration requires actual values of Short, Open and Match standards, which are not trivial at cryogenic temperatures especially for the Match. We characterized the actual values independently using probe~$P_2$, calibrated with the temperature independent TRL algorithm as described before. However, we found that at least for low frequencies~$<$~12~GHz the actual values of the Open, Short, and Match are close enough to ideal corresponding to 1, -1 and 0,  so the latter can be used directly without noticeable difference in the final extracted DUT S-parameters of the probe. This obviously won't be the case at higher frequencies as parasitics become significant.} from \cite{Rytting}, and where square root ERF must be extracted properly using procedures \cite{vna_book}. 

Next, probes $P_1$ and $P_3$ are landed on the thru standard of the calibration wafer, as shown in Fig.~\ref{thru}(b), and the S-parameters $Spa_{11}$, $Spa_{21}$, $Spa_{12}$, and $Sa_{22}$ between calibration planes~4 and~5 (corresponding to ports~1 and~2 of the VNA) are measured. Note that $Sa_{22}$ of the preamplifier network is measured directly, the rest of S-parameters $Sa_{11}$, $Sa_{21}$, $Sa_{12}$, between calibration planes~3 and~5 are obtained by de-embedding the probe~1 $Sp$-parameters using the following equations:

\begin{equation}
   Sa_{11} = \frac{Spa_{11}-Sp_{11}}{Spa_{11}Sp_{22} - Sp_{11}Sp_{22} + Sp_{12}Sp_{21}}\label{eq:sa11}
\end{equation}

\begin{equation}
   Sa_{21} = \frac{Spa_{21}}{Sp_{21}}(1-Sp_{22}Sa_{11})\label{eq:sa21}
\end{equation}

\begin{equation}
   Sa_{12} = \frac{Spa_{12}}{Sp_{12}}(1-Sp_{22}Sa_{11})\label{eq:sa12}
\end{equation}
These $Sa$-parameters are shown in Fig.~\ref{pream_spar}(b). Although we refer to this block as a ``pre-amplifier'' for simplicity, the measurement in fact includes the entire signal chain from probe tip $P_{3}$ through bias-T~27, circulator~28, preamplifier~30, the primary channel of directional coupler~32, and the interconnecting cables up to port~2 of the VNA. 

As such, the extracted $Sa$-parameters represent a comprehensive characterization of both the cryogenic and room-temperature components of the noise receiver front end. For example, the $Sa_{11}$ response exhibits fewer oscillations due to the relatively short electrical path between preamplifier~30 and probe tip $P_{3}$. In contrast, $Sa_{22}$ has a lower magnitude and shows significantly stronger oscillations, which result from the much longer and higher-loss electrical path between the output of the preamplifier~30 and port~2 of the VNA.  


\begin{figure*}[h!]
\begin{multicols}{2}
   \includegraphics[width=3.5in, clip, trim=15pt 15pt 15pt 15pt]{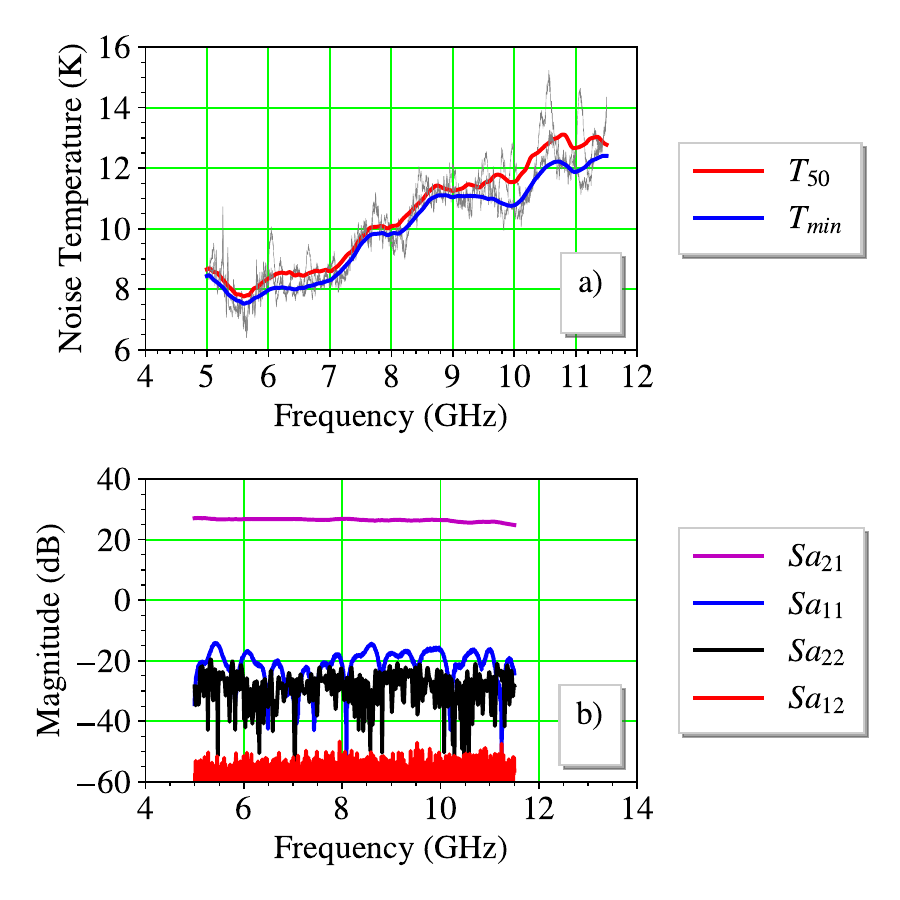}\label{pream_spar} 
   \caption{Measured S-parameters of the preamplifier including probe $P_3$, bias-T~27 and circulator~28. Traces in (a): top: $T_{50}$, bottom: $T_{min}$; traces in (b) top to bottom: $Sa_{21}$, $Sa_{11}$,  $Sa_{22}$,  $Sa_{12}$. }
    \includegraphics[width=3.5in, clip, trim=15pt 15pt 15pt 15pt]{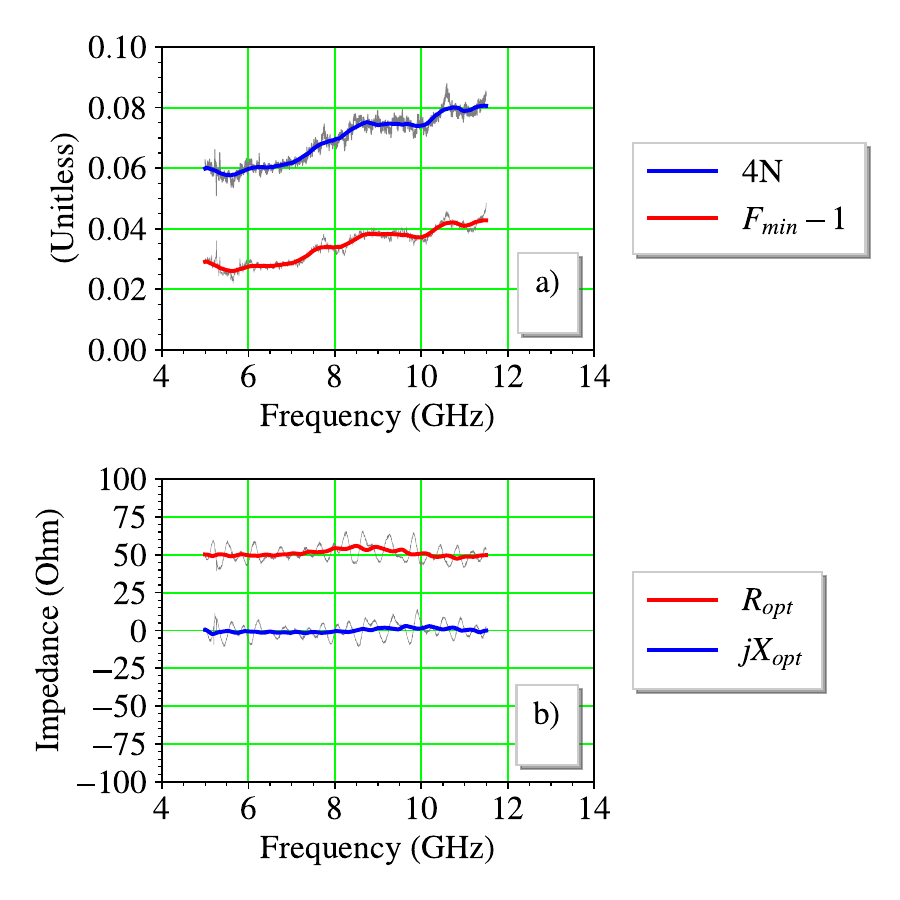}\label{pream_spar}
   \caption{Measured Noise parameters of the preamplifier chain. Traces in (a): top: $4N$, bottom: $F_{min}-1$. Traces in (b): top: $R_{opt}$, bottom: $jX_{opt}$. }\label{preamp_noise_par}
    \end{multicols}
    \vspace{-1.5em}
\end{figure*}

\subsubsection{Noise parameters characterization}
This is the final step of the calibration procedure. Probes $P_3$ and $P_2$ are landed on the thru standard, as shown in Fig.~\ref{thru}(a), and the noise power spectral density (PSD) $W_i$ at port~2 of the VNA is measured for each of the four impedance states $Z_i$ of the tuner. 

As a result of the preceding calibration steps, the following parameters are obtained: the noise PSD $W_i$ corresponding to each tuner state; the $Sa$-parameters of the preamplifier chain from probe tip $P_3$ to VNA port~2; and the tuner impedances $Z_i$. Using this dataset, four mismatch factors $Ma_1$, $Ma_2$, $Ma_3$, and $Ma_4$ are calculated with (\ref{eq:match}), and four corresponding noise factors $Fa_1$, $Fa_2$, $Fa_3$, and $Fa_4$ are obtained using (\ref{eq:noise_factor}), where $T_i = k^{-1} 10^{0.1W_i-3}$. The noise parameters of the preamplifier chain are then extracted using (\ref{eq:noise_param1}). These results\footnote{In all subsequent figures presenting noise parameters, the gray curves represent raw measurement data, while the colored curves show data processed with Savitzky--Golay smoothing using a filter window size of 200, applied to a total of 1001 frequency points.} are shown in Fig.~\ref{preamp_noise_par}. 

The noise temperature of the preamplifier chain, which effectively defines the overall system noise temperature, can then be calculated using (\ref{eq:noise_param}) for any value of DUT output impedance seen by the preamplifier. As an example, the minimum noise temperature $T_{min}$ and the noise temperature for a 50~$\Omega$ source, $T_{50}$, of the preamplifier chain are shown in Fig.~\ref{pream_spar}(a). It can be observed that the resulting noise temperature is significantly higher than the specified noise temperature of the preamplifier alone, which is on the order of 3--4~K \cite{LNF_LNC0_3_14B}. This increase is primarily due to additional loss between probe tip $P_3$ and the preamplifier input, including contributions from the probe itself, the bias-T, the circulator, and the cryogenic interconnecting cables.

At the same time, the difference between $T_{min}$ and $T_{50}$ is negligible due to the high isolation provided by circulator~28, which is the main reason it is used instead of a directional coupler. In fact, this isolation is sufficient to allow the use of only $T_{50}$ for subsequent DUT noise calculations, regardless of the DUT $S_{22}$, which is an important feature of the proposed setup. The impact of this isolation can also be observed in the $R_{opt}$ and $jX_{opt}$ curves shown in Fig.~\ref{preamp_noise_par}(b), where the imaginary part is close to zero and the real part is near 50~$\Omega$.

Finally, as noted in \cite{Belostotski}, Fig.~\ref{preamp_noise_par}(a) shows that the measured minimum noise factor and the Lange invariant satisfy the relationship $F_{min}-1 < 4N$, indicating the absence of fundamental errors in either the measurement setup or the data processing. We found this relationship to be a particularly useful diagnostic tool for debugging the measurement scripts, especially during matrix inversion for ABCD-parameter calculations, which can become computationally intensive when performed over a large number of frequency points.

As a result of the DUT S-parameter calibration and the calibration of the preamplifier S-parameters and noise parameters, once the probes are landed on the actual DUT—either a transistor or an amplifier—the DUT S-parameters $S_{11}$, $S_{21}$, $S_{12}$, and $S_{22}$ can be obtained directly from the VNA using ports~3 and~4. The output noise temperature $T_i$ corresponding to each tuner impedance state $Z_i$, which is required for noise parameter extraction using (\ref{eq:noise_factor}), (\ref{eq:match}), and (\ref{eq:noise_param1}), can be calculated using the following expression (\ref{eq:Ti}):


\begin{equation}
   T_i = \frac{k^{-1} 10^{0.1W_i-3}}{|Sa_{21}|^2 Ma_i} - Ta_i  \label{eq:Ti}
\end{equation}
where $W_i$ is the available absolute noise PSD measured by the noise receiver ”N” 21 of the VNA for each tuner state in units of [dBm/Hz]; $k$ is Boltzmann constant; and $Ta_{DUT}$ is the noise temperature of the preamplifier when loaded with the DUT output impedance calculated using (\ref{eq:noise_param}) and mismatch $Ma_i$ between DUT output and the probe tip $P_3$ is:

\begin{equation}
Ma_i = \frac{1-|\Gamma a_i|^2}{|1-\Gamma a_i Sa_{11}|^2 \Big(1-\Big|Sa_{22} + \frac{Sa_{21}Sa_{12}\Gamma a_i}{1-Sa_{11}\Gamma a_i}\Big|^2 \Big)}  
\end{equation}
where $\Gamma a_i$ is the source reflection coefficient seen by the preamplifier as cascaded tuner ouput reflection and DUT:

\begin{equation}
\Gamma a_i = S_{22} + \frac{S_{21}S_{12}\Gamma_i}{1-S_{11}\Gamma_i} 
\end{equation}

In practice, because of good circulator isolation and matching, $Sa_{11} \approx Sa_{12} \approx Sa_{22} \approx 0$ and therefore $Ma_i$ coefficient significantly simplifies: $Ma_i \approx 1- |\Gamma a_i|^2 $

\begin{figure*}[!t]
\begin{multicols}{2}
   \includegraphics[width=3.5in, clip, trim=15pt 15pt 15pt 15pt]{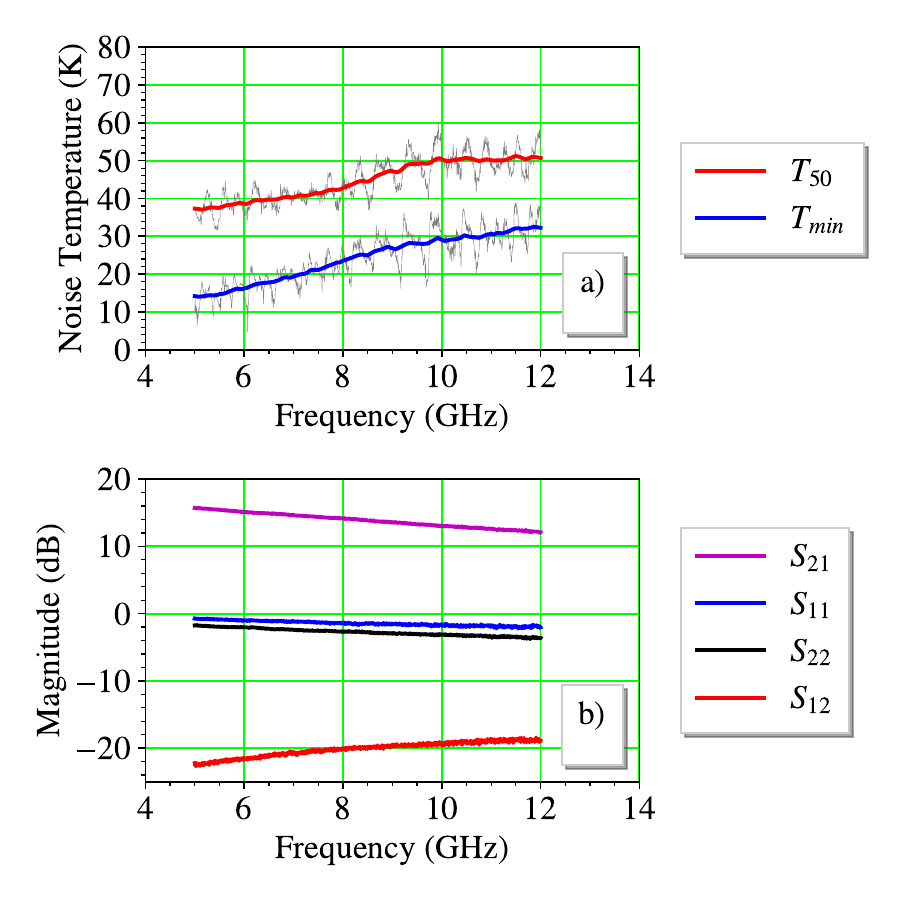}\\
   \caption{FET noise temperatures (a): top: $T_{50}$, bottom: $T_{min}$ and S-parameters (b) top to bottom: $Sa_{21}$, $Sa_{11}$,  $Sa_{22}$,  $Sa_{12}$.}\label{fet_noise_t}

    \includegraphics[width=3.5in, clip, trim=15pt 15pt 15pt 15pt]{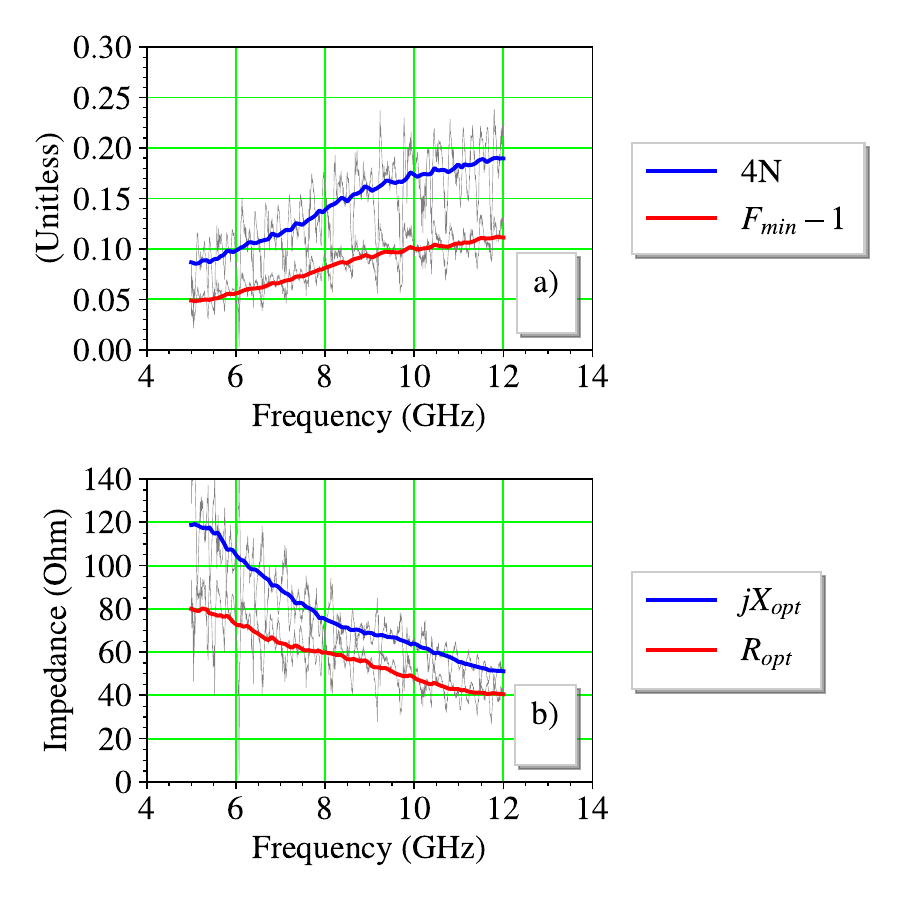}\\
  \caption{Noise parameters of the FET: (a) Noise Factor (bottom) and Lange invariant (top), showing proper inequality $4N>F_{min}-1$, see \cite{Boglione}; (b) optimal resistance (bottom) and reactance of the source (top).}\label{fet_noise_par}
    \end{multicols}
    \vspace{-1.5em}
\end{figure*}
\section{Characterization of Cryogenic FET}
The FETs and the subsequent low-noise amplifiers were implemented using Samsung’s 14~nm FinFET technology, specifically employing super-low-threshold-voltage (SLVT) devices to maximize the unity-gain frequency ($f_T$) and the maximum oscillation frequency ($f_{max}$). To ensure robust performance across validated thermal ranges, the gate length was optimized to minimize the noise temperature ($T_e$) at room temperature and at $-40^\circ$C. In parallel, the gate width was carefully selected to align the real part of the optimum noise impedance ($Z_{opt}$) with the 50~$\Omega$ impedance circle. This design approach enabled a simplified matching network that simultaneously satisfies noise and power matching requirements.

At the time of the original design, cryogenic device data and dedicated noise-parameter characterization setups were not available. Consequently, the characterization framework described in this paper was later employed to investigate the noise parameters of these devices at $T = 4$~K for the first time.


\begin{figure}
  \begin{center}
  \includegraphics[width=3.5in, clip, trim=15pt 15pt 15pt 15pt]{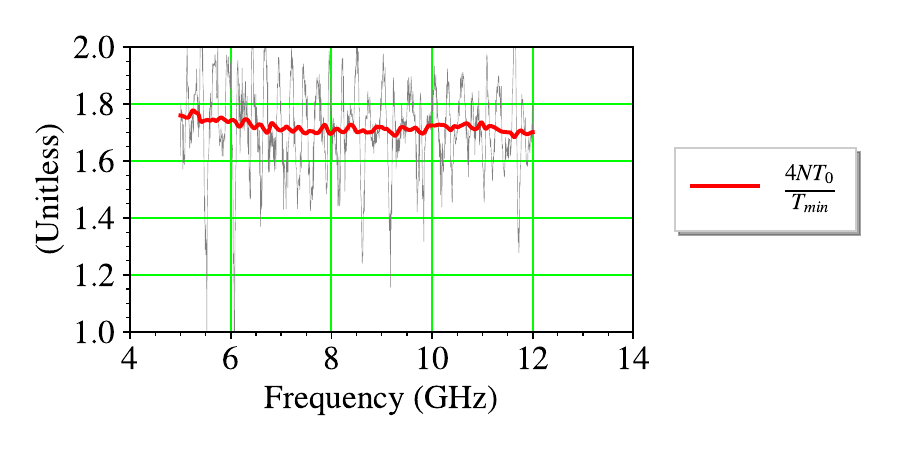}\\
  \caption{The inequality $2>4NT_0/T_{min}>1$ provides a useful check to validate the measured noise parameters.}\label{NTest}
  \end{center}
\end{figure}

\begin{figure}[!h]
  \begin{center}
  \includegraphics[width=3.5in, clip, trim=10pt 10pt 10pt 10pt]{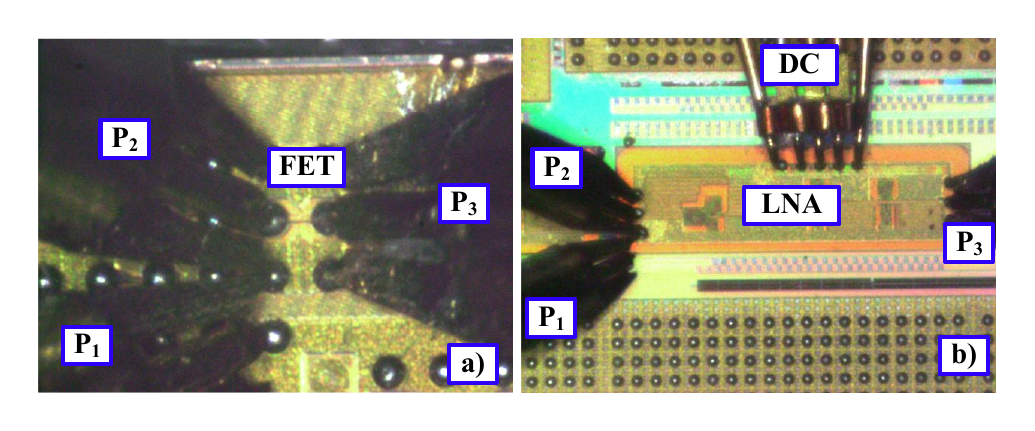}\\
  \caption{Measurement of the FET (a) and LNA based on this FET (b)}\label{fet_lna_m}
  \end{center}
\end{figure}
The FET S-parameters and noise temperature measurement results are shown in the fig.\ref{fet_noise_t}, noise parameters are shown in fig.\ref{fet_noise_par}. An important result that validates the calibration of the setup is shown in Fig.~\ref{NTest}. As previously discussed in \cite{phdt} for both FETs and bipolar transistors, a necessary condition for a physically valid set of noise parameters is that the ratio $4NT_0/T_{min}$ lies between 1 and~2. Fig.~\ref{fet_lna_m} shows microscope view of a wafer with individual FET (a) and LNA (b) being measured. 


\begin{figure*}[t!]
\begin{multicols}{2}
   \includegraphics[width=3.5in, clip, trim=15pt 15pt 15pt 15pt]{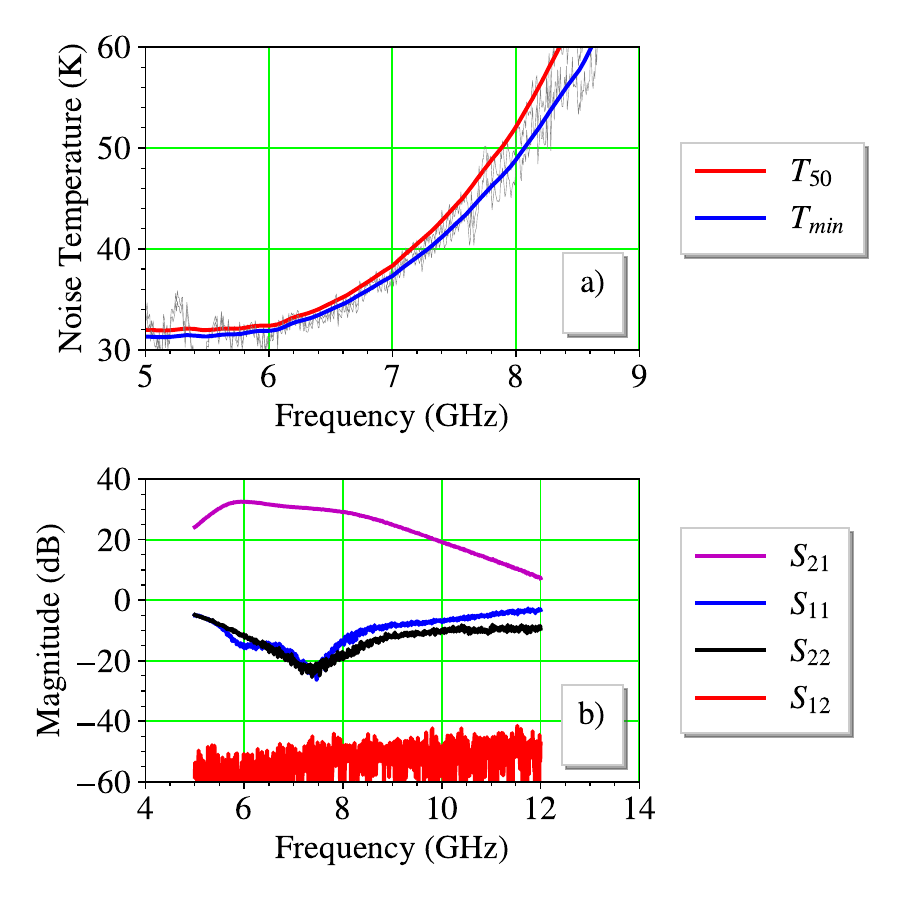}\\
  \caption{LNA noise temperatures (a): top: $T_{50}$, bottom: $T_{min}$ and S-parameters (b) top to bottom: $Sa_{21}$, $Sa_{11}$,  $Sa_{22}$,  $Sa_{12}$.}\label{lna_spar}
   
    \includegraphics[width=3.5in, clip, trim=15pt 15pt 15pt 15pt]{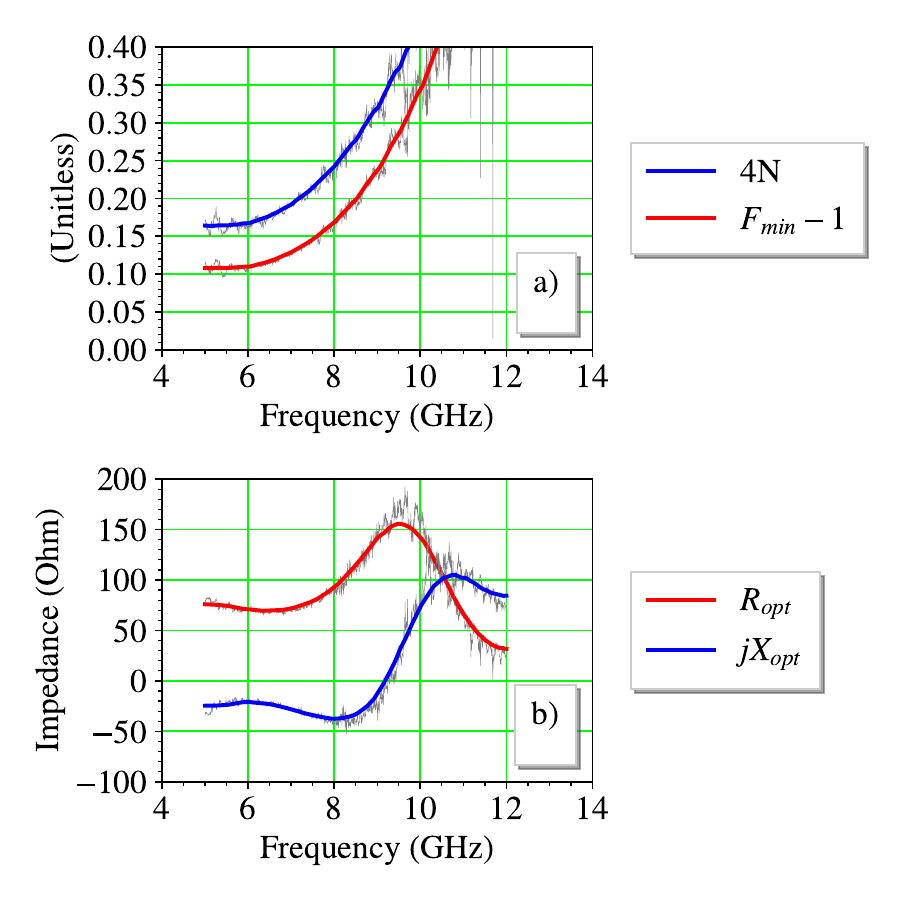}\\
   \caption{Noise parameters of the LNA: (a) Noise Factor (bottom) and Lange invariant (top); (b) optimal resistance (top) and reactance of the source (bottom).}\label{lna_noise_par}
    \end{multicols}
     \vspace{-2em}
\end{figure*}

It is important to note the following observation. As can be seen in Fig.~\ref{fet_noise_t}, the average $S_{21}$ of the FET is approximately 13~dB over the 5--12~GHz frequency range. For a DUT operating at $T = 4$~K, the thermal noise floor is~$-193$~dBm/Hz. Even for a DUT with a low noise figure and a gain of $S_{21} = 15$~dB, the resulting output noise PSD is only~$-178$~dBm/Hz, which remains below the room-temperature instrumentation noise floor of~$-174$~dBm/Hz. Consequently, direct noise temperature measurements are not possible using conventional methods.
In the proposed setup, however, this limitation on DUT gain is alleviated by the inclusion of a rigorously calibrated preamplification chain with well-characterized gain and noise temperature. As a result, reliable noise parameter characterization of individual FETs is possible even for devices with relatively low gain.

\section{Characterization of cryogenic LNA}
\begin{figure}[h]
  \includegraphics[width=3.5in, clip, trim=40pt 80pt 60pt 90pt]{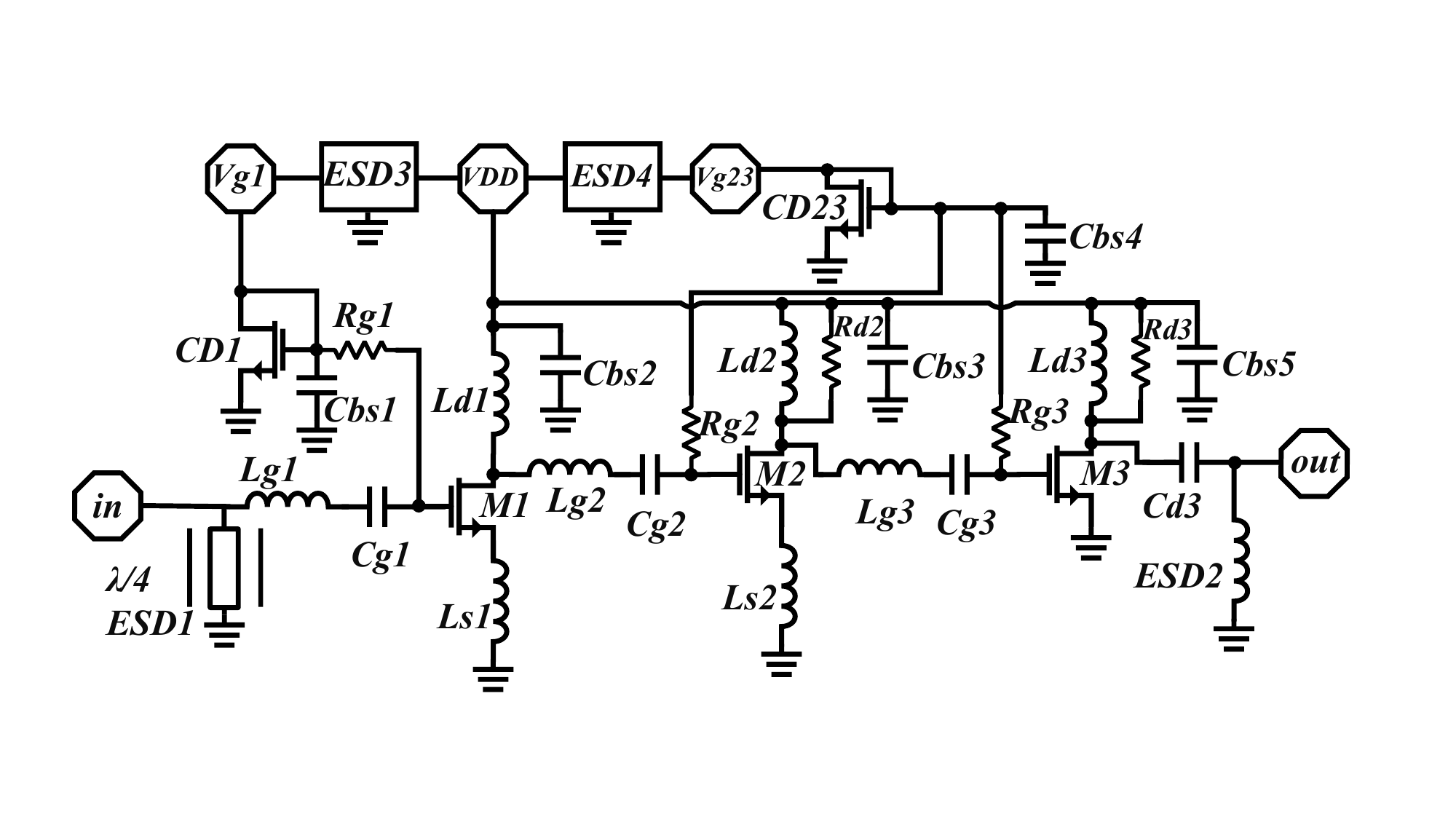}\\
  \caption{3-stage common-source LNA schematic, the "in", "out", "Vg1", "VDD", and "Vg23" octagons correspond to the pads on the die for probe landing.}\label{LNA_schematic}
\end{figure}

The schematic of the LNA is shown in Fig.~\ref{LNA_schematic}. The amplifier consists of three cascaded common-source stages and is implemented in a 14~nm FinFET technology. The first stage is input-matched to 50~$\Omega$ by resonating the input capacitance of NFET~M1 with inductors $L_{g1}$ and $L_{s1}$ at the 7~GHz center frequency. In parallel, the width of M1 is selected such that the real part of the input impedance is 50~$\Omega$. This approach enables simultaneous matching of the input impedance and the optimum noise impedance to 50~$\Omega$. The same design technique is applied to the second stage to further improve the overall noise performance of the LNA.

One drawback of the common-source architecture is limited stage-to-stage isolation, which can potentially lead to instability. However, this reduced isolation also enables the introduction of resonances in subsequent stages, which can be exploited to achieve broadband input matching. The drains of transistors M1, M2, and M3 are loaded with inductors $L_{d1}$, $L_{d2}$, and $L_{d3}$, respectively. The quality factors of $L_{d2}$ and $L_{d3}$ are intentionally reduced by incorporating resistors $R_{d2}$ and $R_{d3}$ to enhance LNA stability. Interstage matching is realized using $L_{d1}$, $L_{d2}$, $L_{g2}$, $L_{g3}$, $C_{g2}$, $C_{g3}$, and $R_{d2}$. The LNA output is matched to 50~$\Omega$ using $L_{d3}$, $R_{d3}$, $C_{d3}$, and inductor ${\mathrm{ESD2}}$. All inductors are implemented by stacking the top two metal layers in order to minimize loss.

The drain supply voltage applied at the VDD pad typically ranges from 0.3 to 0.7~V. The drain current of the first stage is controlled by injecting a current bias at the Vg1 pad into the catch diode CD1, which mirrors the current into the drain of M1. Similarly, CD23 mirrors the current injected at the Vg23 pad into the M2 and M3 stages. Independent current biasing of the first stage and stages 2 and 3 provides additional flexibility for trading off noise, power consumption, and gain.

The LNA is protected against electrostatic discharge (ESD) through a combination of protection elements, including a quarter-wavelength transmission line ESD1 connected to ground, foundry-standard ESD diodes ESD3 and ESD4, and a custom RF inductor ESD2 connected to ground. These protection structures are implemented at the input, Vg1, VDD, Vg23, and output pads, respectively.

Measurement results of the LNA based on the FETs described above, obtained at $T=4$~K, are shown in Figs.~\ref{lna_spar} and~\ref{lna_noise_par}.

\section{Discussion of Results}
Uncertainty analyses for tuner-based cryogenic noise measurements of connectorized DUTs have been investigated in detail in prior work \cite{Bel_big_paper}. It was shown in \cite{bel_arftg_paper} that the uncertainty of the noise temperature measurement for each tuner state is primarily dominated by uncertainty in the physical temperature of the tuner and by uncertainty in the DUT gain. Our results are consistent with these earlier observations, as can be inferred from (\ref{eq:noise_factor}) and from the trace noise behavior of a well-matched preamplifier in Fig.~\ref{preamp_noise_par}(a), a poorly matched FET in Fig.~\ref{fet_noise_t}(a), and a moderately matched LNA in Fig.~\ref{lna_spar}(a).

For example, Fig.~\ref{lna_spar} shows that the frequency region around 7.5~GHz exhibits the best input matching. Correspondingly, the noise temperature and noise factor curves in this region demonstrate minimal trace noise, on the order of 1~K. In contrast, in frequency regions where the input matching degrades to worse than $-10$~dB, the trace noise increases and approaches approximately 5~K. In \cite{bel_arftg_paper}, a Monte Carlo analysis was used to estimate uncertainties in the extracted noise parameters $T_{min}$, $N$, and $\Gamma_{opt}$. In the context of the present setup and calibration scheme, applying a similar approach would require accurate knowledge of the uncertainties associated with cryogenic on-wafer TRL calibration, which remains an active area of research.

\begin{figure}
  \begin{center}
  \includegraphics[width=3.5in, clip, trim=10pt 10pt 10pt 10pt]{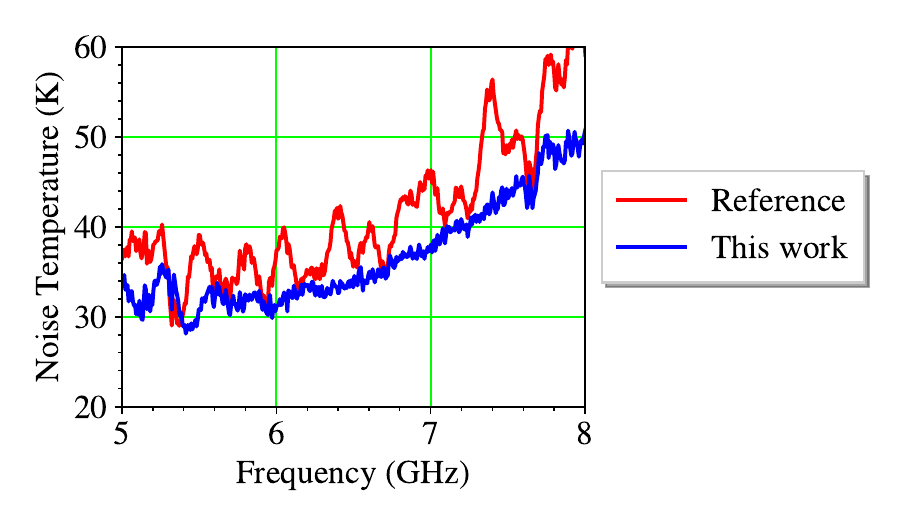}\\
  \caption{LNA noise temperature measured by the setup in this paper (bottom trace) and by the scalar on-wafer method from \cite{arftg_paper} (top trace), the latter trace noise is in agreement with $\pm 6$ K uncertainty demonstrated in \cite{arftg_paper} for input match of -10...-15 dB.}\label{lna_comp}
  \end{center}
\end{figure}

Instead, we employed an alternative validation approach. To verify the noise temperature results, an independent measurement of the same LNA was performed using a purely scalar on-wafer method previously described in \cite{arftg_paper}. This reference technique enables the measurement of scalar gain and noise temperature at a source output impedance of approximately $Z \approx 50~\Omega$. A comparison of the noise temperature results obtained using both methods is shown in Fig.~\ref{lna_comp}, where the traces from the two measurement techniques are overlaid.

The uncertainty of the scalar reference method depends on the input matching of the DUT, as discussed in \cite{arftg_paper}. For the LNA presented in this work, with $S_{11}$ in the range of $-10$ to $-15$~dB, the corresponding uncertainty is on the order of 5--6.5~K. As shown in Fig.~\ref{lna_comp}, the results obtained using the reference scalar method are in close agreement with those derived from the proposed noise parameter measurement setup, thereby validating the presented approach. 

Unlike the LNA, the FET cannot be characterized using the method described in \cite{arftg_paper} due to its significantly lower gain. However, the quality of the cryogenic calibration for FET measurements can be demonstrated using the following example. 

\begin{figure}[!h]
  \begin{center}
  \includegraphics[width=3.5in, clip, trim=10pt 10pt 10pt 10pt]{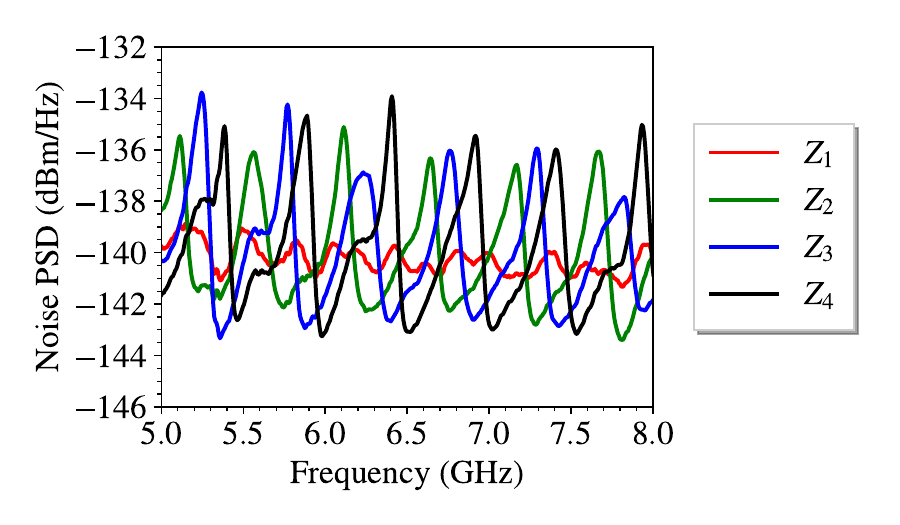}\\
  \caption{Raw noise power spectral density produced by the FET and preamplifier in 4 impedance states of the tuner, as measured by the noise receiver. If counting traces from left to right, trace whose peak is the 1st corresponds to $Z_2$, 2nd: $Z_3$, 3rd: $Z_4$, remaining trace in the center: $Z_1$.}\label{psd_raw}
  \end{center}
\end{figure}

Figure~\ref{psd_raw} shows the raw noise PSD of the FET as measured by the VNA noise receiver for each of the four tuner impedance states. It can be observed that, in states $Z_2$, $Z_3$, and $Z_4$, the PSD traces oscillate between approximately $-142$ and $-136$~dBm/Hz. These oscillations arise from impedance mismatch between the tuner and the FET, combined with the presence of a long cryogenic wire connecting the tuner to probe $P_2$, which causes variation in the available gain of the FET as frequency changes. At the same time, the extracted $F_{min}$ curve shown in Fig.~\ref{fet_noise_par}(a) exhibits a trace noise of only 0.05 units, corresponding to a noise figure of $\mathrm{NF} = 10\log(1.05) \approx 0.2$~dB. The fact that four raw PSD traces with a magnitude variation of approximately 6~dB are processed to yield a final de-embedded noise figure with a variation of only $\pm 0.1$~dB demonstrates the high precision of the proposed calibration procedure and its ability to effectively remove systematic measurement errors.

It is likely that even better results and lower trace noise could be achieved in the future if the following improvements are implemented:
\begin{enumerate}
\item instead of using the commercial CS-105 TRL calibration substrate employed in this work, a custom TRL calibration kit fabricated on the same wafer as the DUT could be designed, thereby eliminating potential impedance mismatches between the test wafer and the separate calibration substrate;
\item the tuner is moved closer to the DUT or even integrated directly into the probe assembly, thereby reducing the number of standing waves between the DUT input and the source.
\end{enumerate}

\section{Conclusion}
In summary, this paper presents a measurement approach for direct \textit{on-wafer} characterization of noise parameters of DUTs at $T = 4$~K in the C- and X-bands. The measurements are performed using the cold-source method in combination with a source-pull technique implemented with a four-state cryogenic tuner. A calibration scheme is described that enables vector-corrected S-parameter and noise parameter measurements to be carried out within a single cryostat cooldown. The methodology is demonstrated through noise parameter measurements of a single on-wafer FET and an LNA realized using this device. The LNA noise temperature results are independently verified through comparison with a previously reported scalar on-wafer measurement technique, showing close agreement over the measured frequency range. The presented setup and calibration procedure illustrate a practical pathway toward simultaneous \textit{on-wafer} S-parameter and noise parameter measurements of semiconductor devices at cryogenic temperatures.
\newpage
\section*{Acknowledgment}

The authors would like to thank Michael Himmelfarb and Leo Belostotski for the initial help with tuner operation, as well as Steve Dudkiewicz and Diogo Ribeiro of Maury Microwave for providing MATLAB examples of connectorized amplifier measurements. We also want to thank Kevin Tien for the suggestion to use tuners for IBM cryo LNA project and Brian Gaucher for his support of this idea, John Timmerwilke and Ray Robertazzi for help with the cryogenic and vacuum systems.  Finally, we thank Alberto Valdes Garcia for general project management support. Special thanks to Jalina Vyrva for help in the manuscript preparation.

\ifCLASSOPTIONcaptionsoff
  \newpage
\fi



\bibliographystyle{ieeetr}

\vfill


\end{document}